\documentclass[reprint, aip, apl, amsmath, amssymb]{revtex4-1}

\usepackage{graphicx}	
\usepackage{dcolumn}   	
\usepackage{bm}		
\usepackage{epstopdf}

\begin{document}

\title{Edge Trapping of Exciton-Polariton Condensates in Etched Pillars}

\author{D. M. Myers}
\email{dmm154@pitt.edu}
\author{J. K. Wuenschell}
\altaffiliation[J. K. Wuenschell is now at ]{Physical Sciences Laboratory, The Aerospace Corporation, El Segundo, CA 90245, USA}
\author{B. Ozden}
\author{J. Beaumariage}
\author{D. W. Snoke}
\affiliation{Department of Physics \& Astronomy, University of Pittsburgh, 3941 O'Hara Street, Pittsburgh, Pennsylvania 15260, USA}
\author{L. Pfeiffer}
\author{K. West}
\affiliation{Department of Electrical Engineering, Princeton University, Princeton, New Jersey 08544, USA}

\date{\today}

\begin{abstract}
In this letter, we present a study of the condensation of exciton-polaritons in large etched pillar structures that exhibit shallow edge trapping. The $\approx100\:\mu$m$\:\times100\:\mu$m pillars were fabricated using photolithography and a BCl$_3$/Cl$_2$ reactive ion etch. A low energy region emerged along the etched edge, with the minima $\approx7\:\mu$m from the outer edge.  The depth of the trap was $0.5-1.5\:$meV relative to the level central region, with the deepest trapping at the corners. We were able to produce a Bose-Einstein condensate in the trap near the edges and corners by pumping non-resonantly in the middle of the pillar. This condensate began as a set of disconnected condensates at various points along the edges, but then became a single mono-energetic condensate as the polariton density was increased. Similar edge traps could be used to produce shallow 1D traps along edges or other more complex traps using various etch geometries and scales.

\end{abstract}

\maketitle


In the past two decades, many experiments have used polaritons resulting from strong coupling between trapped microcavity photons and quantum well (QW) excitons. These bosonic particles have a very light mass ($\sim10^{-4}m_\mathrm{e}$) due to being partially photonic, but also strong particle-particle interactions from being partially excitonic.\cite{Deng2010} This combination of a light mass and strong interactions leads to the formation of Bose-Einstein condensates (BECs) at relatively high temperatures ($\sim10\:$K).\cite{Deng2002Science, Kasprzak2006, Balili2007} Polaritons provide a promising system for studying bosonic particles at even higher temperatures, and polariton lasing has been observed at room temperature in both GaN\cite{Christopoulos2007} and organic\cite{ Kena2010}  systems.


Many methods of confinement have been used to study polariton dynamics in a variety of geometries. Applying stress to a thin ($\approx100\:\mu$m) GaAs sample can be used to shift the exciton energy, resulting in a harmonic trap.\cite{Balili2007, Nelsen2009} Pumping such a stress trap non-resonantly in the center forms a repulsive barrier and can be used to form a ring geometry.\cite{Liu2015} Complex pumping geometries can also be used to confine polaritons, including the use of two or more pump spots in various arrangements or using a ring-shaped pump spot.\cite{Tosi2012, Cristofolini2013, Askitopoulos2013, Askitopoulos2015, Sun2017} More permanent methods of confinement include producing a spacer in certain regions of the cavity during the growth process,\cite{Daif2006, Kaitouni2006, Winkler2015} using sub-wavelength gratings as the top mirror,\cite{Zhang2014, Zhang2015} depositing metal strips onto the top mirror,\cite{Lai2007} and etching the sample after growth to form 1D wires, 2D pillars, and 2D arrays of coupled pillars.\cite{Gutbrod1998, Bajoni2008, Wertz2010, Anton2012, Anton2013, Jacqmin2014}

While optically induced trapping potentials have the advantage of being easily reconfigured, etched trapping allows the confinement to be somewhat independent of the pump laser. Post-growth etching also produces much higher potential barriers at the etched edges than the deposition of metal strips, and it is compatible with our existing sample materials and growth methods, unlike sub-wavelength gratings or modulating the cavity spacing mid-growth. In this letter, we present results from etched pillars similar to many of those mentioned above, but generally larger in scale ($\approx100\;\mu$m$\:\times100\;\mu$m). The pillars were fabricated with a relatively simple photolithography process and alternative BCl$_3$/Cl$_2$ reactive ion etch (RIE), as opposed to the more complicated electron beam lithography and HBr RIE used elsewhere.\cite{Bajoni2008, Wertz2010}  Our pillars exhibit a reduced lower polariton energy near the edges of the pillar independent of the repulsive potential at the location of the pump spot, which has been observed in etched structures similar in scale to the ones presented here.\cite{Anton2012, Anton2013}  In this paper we describe the fabrication of these edge traps, the potential-energy landscape, and the formation of a BEC within that landscape.

\begin{figure}
\centering
\includegraphics{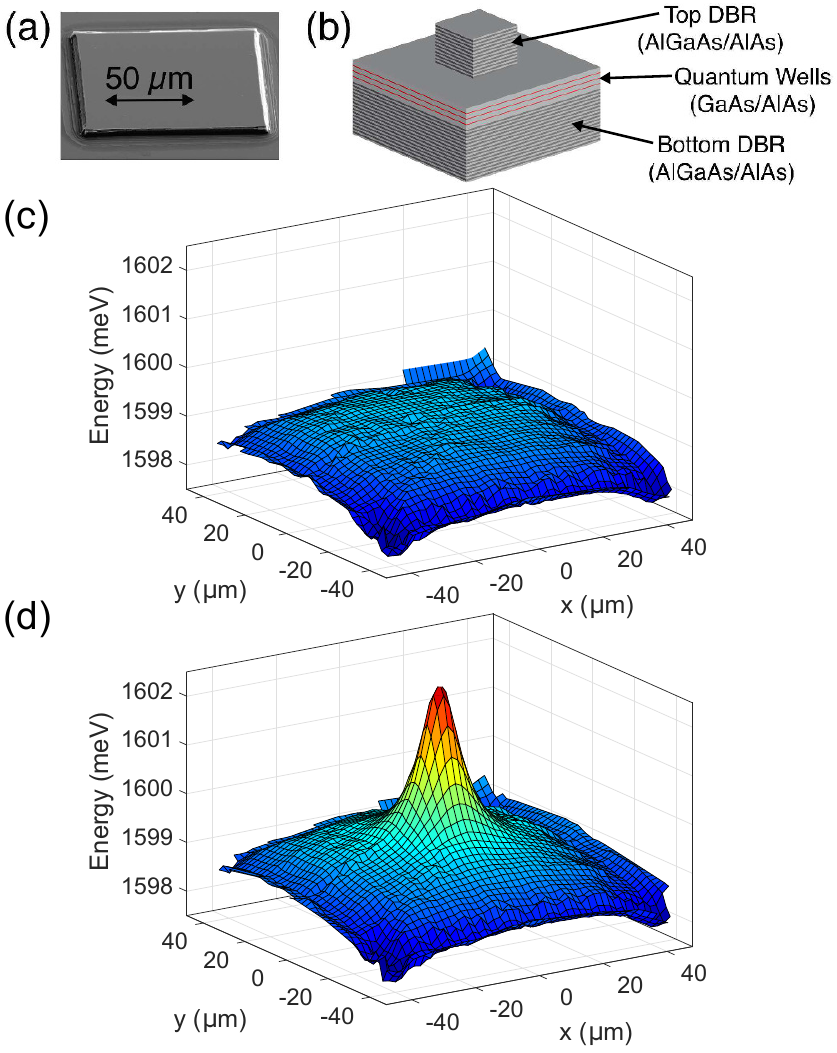}
\caption{(a) SEM image of an etched pillar. (b) The etched pillar structure, with the top DBR removed except in a square area. The bottom DBR and QW layers are un-etched, leaving an open cavity except at the location of the pillar. (c) Linearly interpolated data showing the resulting low density potential landscape. A mostly flat region in the middle gives way to low energy regions at the edges, with minima at the corners. The overall tilt of the landscape is due to the sample gradient. (d) The same surface shown in (c) with a fit of the effective potential from a focused cw pump laser at 51\:mW  added in the center.}
\label{diagram}
\end{figure}

The microcavity samples used for this study were very similar to those used in previous work.\cite{Nelsen2013, Steger2013, Steger2015, Liu2015} They were grown on a GaAs substrate using molecular beam epitaxy (MBE). A $3\lambda/2$ cavity was formed by two Al$_{0.2}$Ga$_{0.8}$As/AlAs distributed Bragg reflectors (DBRs) with 32 periods in the top DBR and 40 periods in the bottom DBR. Within the cavity were three sets of four coupled GaAs/AlAs QWs, with one set at each antinode of the cavity photon mode. These samples were then etched to a depth of  $\approx4\:\mu$m using a $\approx2.7\:\mu$m photoresist mask. The etching was done with a 20:7 BCl$_3$/Cl$_2$ inductively coupled plasma (ICP) reactive ion etch (RIE) at $3.0\:$mT chamber pressure, $600\:$W ICP power, and $75\:$W RF bias power. This removed the top DBR to form  $\approx100\:\mu$m$\:\times100\:\mu$m pillars (Figure \ref{diagram}(a-b)). The lower polariton energy at resonance is $\approx 1601$ meV, and the pillars used in this study were at slightly photonic detuning ($\delta = E_\mathrm{cav}-E_\mathrm{ex} \approx -3\:$meV). 

Except where noted, the pillar was pumped non-resonantly ($E_\mathrm{pump} =1754\:$meV) at an incident angle of approximately $18^{\circ}$ with a continuous wave (cw) Ti:sapphire pump laser. The sample was kept at constant temperature in a liquid helium cryostat ($\approx5\:$K). In order to prevent heating, the pump was mechanically chopped with a 1.3\% duty cycle at 400\:Hz. All reported powers are the peak power of each short pulse. The pump spot was $\approx20\:\mu$m FWHM and centered on the pillar. Measurements were taken using a charge-coupled device (CCD) camera on the output of a spectrometer to allow for energy-resolved imaging. We also obtained in-plane momentum information using angle-resolved imaging. Except in the case of the angle-resolved imaging, the collection angle was limited to $\approx4^{\circ}$ from normal emission with an aperture at the Fourier plane of the objective lens.


The potential landscape of these pillars at low density was acquired using a defocused pump and collecting energy-resolved slices of the pillar at many locations. The interpolated results are shown in Figure \ref{diagram}(c), showing the overall effect of the etch process on the ground state energy of the lower polaritons. A low energy region is apparent at all of the edges, with the minima $\approx7\:\mu$m in from the outer etched edge, and overall energy minima appearing at the corners. The depth relative to the flat region in the middle is $0.5-1.5\:$meV.  This is much shallower than the trapping attained from the smaller scale etching or the cavity spacing methods described above, but comparable to the depths attained from stress traps. The overall tilt shows the direction of the cavity gradient (approximately $75^\circ$ from the $+$x-axis) resulting from a wedge in the cavity thickness.\cite{Balili2006} The lifetime of the polaritons was not affected by the etch process to a degree detectable in the energy linewidth using our equipment. As mentioned elsewhere,\cite{Steger2013} our spectral resolution (0.1~meV) gives a lower bound on the polariton lifetime that is much smaller than the measured lifetimes around 200 ps.\cite{Steger2015}

\begin{figure}
\centering
\includegraphics{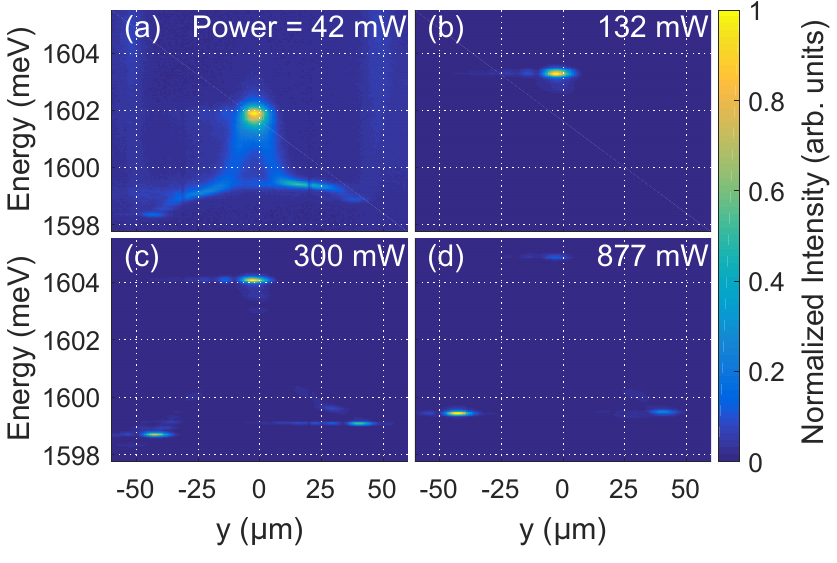}
\caption{Energy-resolved polariton luminescence intensity near $k_\parallel=0$ for a slice across the center of the pillar (along the y-axis defined in Figure \ref{diagram}(c)). The peak pump power (the power of each quasi-cw pulse) for each image was (a) 42 mW, (b) 132 mW, (c) 300 mW, and (d) 877 mW. The vertical streaks at both edges in (a) are from bare exciton luminescence. The horizontal streaks near bright regions are an artifact of the imaging optics. The intensity is normalized separately for each image, so the values are not comparable between images.}
\label{EvR}
\end{figure}

The potential landscape changed with the addition of a focused pump. The repulsive exciton-exciton interactions formed a new effective potential, with an energy ``hill'' which is highest at the location of the pump spot (Figures \ref{diagram}(d) and \ref{EvR}(a)).\cite{Nelsen2013, Wertz2010} This hill became higher at high pump power due to the increase in the exciton density. Figure \ref{EvR} shows the polariton luminescence of a real space slice across the middle of the pillar containing the pump spot, which is apparent in the center, at several powers. This particular slice follows the y-axis and intersects the origin, using the axes defined in Figure \ref{diagram}(c). At low pump power, the polariton luminescence was clearly visible at all points across the pillar, but the highest density was at the pump spot (Figure \ref{EvR}(a)). The pump spot was even more dominant at higher powers (Figure \ref{EvR}(b)). The density distribution underwent a dramatic shift at even higher powers, with the edge regions emitting comparable intensities to the central pump spot (Figure \ref{EvR}(c)). The narrow spectral width of the emission from these regions indicates that they contained Bose condensates. This is also consistent with the dramatically increased transport distance, which can be associated with superfluidity that allowed the polaritons to find the local minima of the potential landscape. At the highest powers, the condensates in the edge regions were blue-shifted to higher energies, eventually becoming equal in energy (Figure \ref{EvR}(d)), consistent with a phase-locked single condensate.

\begin{figure}
\centering
\includegraphics{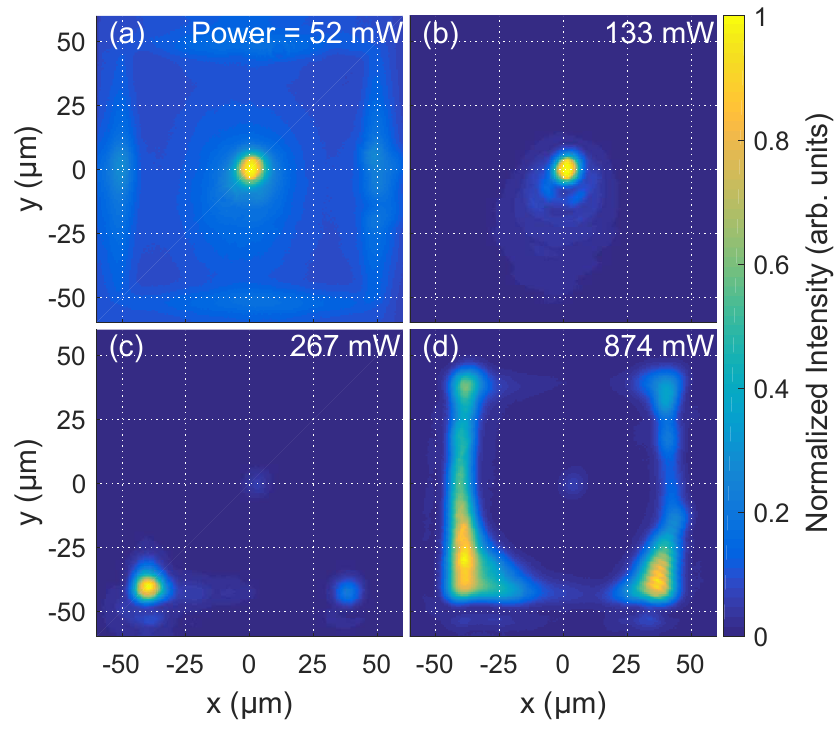}
\caption{Full real space polariton luminescence intensity near $k_\parallel=0$ at  peak pump powers (the power of each quasi-cw pulse) of (a) 52mW, (b) 133mW, (c) 267mW, and (d) 874mW. The coordinate axes are the same as those defined in Figure \ref{diagram}(c). The intensity was normalized separately for each image, so the values are not comparable between images.}
\label{Rspace}
\end{figure}

A similar story emerges from the full real-space imaging of the entire pillar at various pump powers (Figure \ref{Rspace}). The pump spot was prominent at low powers, but polariton luminescence was visible across the whole pillar (Figure \ref{Rspace}(a)). The relatively bright luminescence at the edges and around the pillar was from the bare excitons in the open cavity. As the power was increased (Figure \ref{Rspace}(b)) the pump spot became even more dominant, but the polaritons remained in the non-condensate regime and their diffusion length remained low. At higher powers (Figure \ref{Rspace}(c)), the corners became the dominant features as polaritons flowed long distances to fill in the locations of the energy minima. As the power was increased further (Figure \ref{Rspace}(d)), the polaritons filled a connected region extending along all the edges. The asymmetry relative to the pillar shape was caused by the cavity gradient, with the side closest to $\mathrm{y}=-50\:\mu$m at lower overall energy, and the lower left corner ($\mathrm{x}=-50\:\mu$m) at the lowest point.

\begin{figure}
\centering
\includegraphics{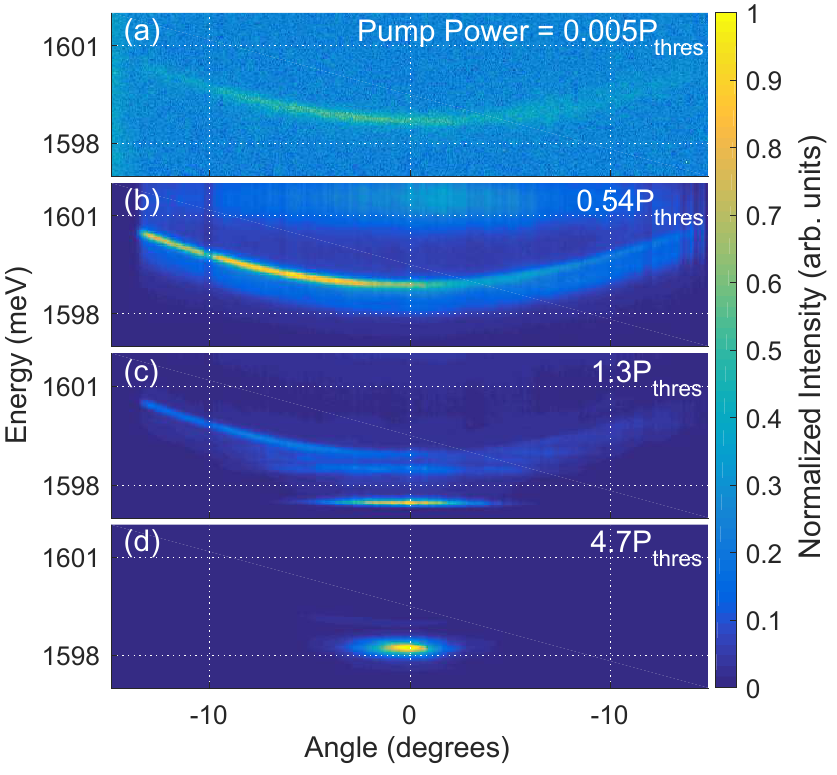}
\caption{Energy vs. angle of emission at various pump powers. The collection angle was along the direction close to the x-axis defined in Figure \ref{diagram}(c), but the luminescence was collected from the entire pillar. The pillar was pumped  at $\approx1759\:$meV and at normal incidence, with a threshold power ($P_{\mathrm{thres}}$) of $\approx86\:$mW, as defined similarly to those in Figure \ref{IvP}. The pump powers are (a) 0.005$P_{\mathrm{thres}}$, (b) 0.54$P_{\mathrm{thres}}$, (c) 1.3$P_{\mathrm{thres}}$, and (d) 4.7$P_{\mathrm{thres}}$. The intensity was normalized separately for each image, so the values are not comparable between images. }
\label{KspaceSeries}
\end{figure}

Figure \ref{KspaceSeries} shows the angle-resolved polariton luminescence, with the measured angle corresponding to momentum along the direction close to x-axis defined in Figure \ref{diagram}(c). This data was collected from a different but nearly identical pillar (in terms of etch parameters, size, detuning, and source sample). The pump in this case was at $\approx1759\:$meV and at normal incidence. This resulted in considerably different power levels for the same luminescence characteristics, but the overall characteristics were the same. At low pump power (Figure \ref{KspaceSeries}(a)) the polaritons were emitted at many angles fairly evenly, corresponding to the filling of many momentum states. At higher power, high-energy emission emerged from the top of the repulsive ``hill'' at the pump spot, but the dispersion remained generally the same. Above the condensate threshold power (Figure \ref{KspaceSeries}(c)) the polaritons moved overwhelmingly into a narrow low energy state near $k_\parallel=0$, which is characteristic of Bose-Einstein condensation. At even higher pump powers (Figure \ref{KspaceSeries}(d)) the parabolic polariton dispersion became nearly undetectable compared to the luminescence at  $k_\parallel=0$, indicating a large condensate fraction. The condensate also exhibited a slight blue shift in energy due to the high polariton density along the edges of the pillar.

\begin{figure}
\centering
\includegraphics{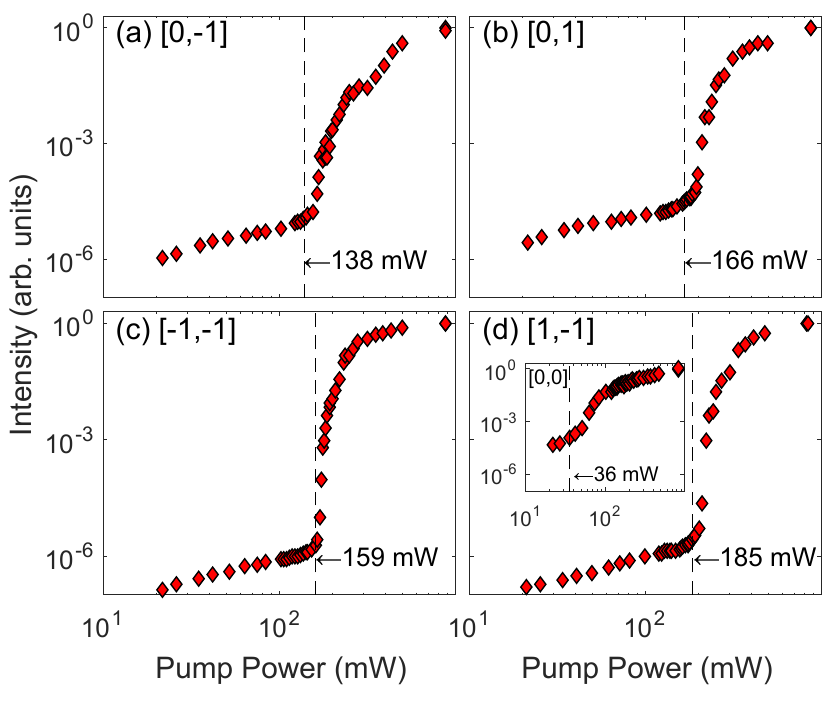}
\caption{Log-log plot of the intensity of the polariton luminescence at the pillar edges vs. the pump power at various positions. Intensity values were taken by integrating over a small area in spectrally resolved images at the lower polariton energy at the pillar edge in images similar to those in Figure \ref{EvR}. The vertical lines mark the threshold powers, which are defined by the onset of nonlinearity. Positions are defined according to Figure \ref{thresholds}(b). The threshold powers and positions for each plot are (a) $P_{\mathrm{thres}}=138\:$mW at point [0, -1], (b)  $P_{\mathrm{thres}}=166\:$mW at point [0, 1], (c) $P_{\mathrm{thres}}=159\:$mW at point [-1, -1], (d)  $P_{\mathrm{thres}}=185\:$mW at point [1, -1], and (inset)  $P_{\mathrm{thres}}=36\:$mW at point [0, 0].}
\label{IvP}
\end{figure}

\begin{figure}
\centering
\includegraphics{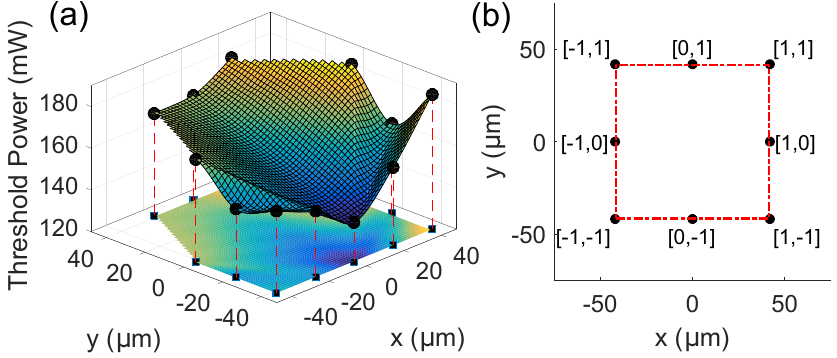}
\caption{(a) Threshold powers (as defined in Figure \ref{IvP}) at various points on the pillar. The black circles give the actual threshold powers at each point. The surface is there to help guide the eye, and is a linear interpolation of the data points. The axes are the same as those in Figure \ref{diagram}(c). (b) Diagram defining a simple designation for each point, given as the actual position in Figure \ref{diagram}(c), but in units of the shortest distance from the center to the bottom of the edge trap ($\approx41\:\mu$m).}
\label{thresholds}
\end{figure}

To determine the precise threshold power, we plotted the polariton luminescence intensity vs. power at various points along the pillar (Figure \ref{IvP}). The positions of each of these points are defined using the same coordinate axes as Figure \ref{diagram}(c), but in units of the shortest distance from the origin to the edge trap ($\approx41\:\mu$m) (see Figure \ref{thresholds}(b)). At each position, a linear dependence on intensity was maintained up to the condensate threshold, where a strong nonlinearity emerged. The threshold power was generally lower along the edges (Figure \ref{IvP}(a-b)) than at the corners  (Figure \ref{IvP}(c-d)), with the lowest threshold on the middle of the edge downhill along the gradient from the pump spot, consistent with the picture that the majority of polaritons flow downhill (see also Figure \ref{thresholds}(a)). We note that the increase in emission from the edges was more than five orders of magnitude at the threshold in some cases. The emission eventually saturated and returned to increasing linearly with pump power.

The emission from the pump spot region showed a much less sharp and lower threshold ($\approx36\:$mW) (Figure \ref{IvP}(d)(inset)). As discussed in Ref.~\onlinecite{Nelsen2013}, we interpret this behavior at the pump spot as the onset of a quasi-condensate at densities just below true condensation; the momentum distribution is strongly altered, but there is no evidence of superfluidity. At higher density, a true condensate emerges in whatever local energy minimum is available.

A compilation of the threshold powers at various points along the edge is shown in Figure \ref{thresholds}(a). The line of symmetry and overall tilt is oriented along the sample gradient. In general, higher energy points along the gradient had higher threshold powers, as expected, since polaritons from the pump region will generally run downhill and away from these regions. The corners are the obvious exception, since the two shown had the lowest energies of any points on the pillar, but noticeably higher threshold powers than the neighboring points on the edges.

\begin{figure}
\centering
\includegraphics{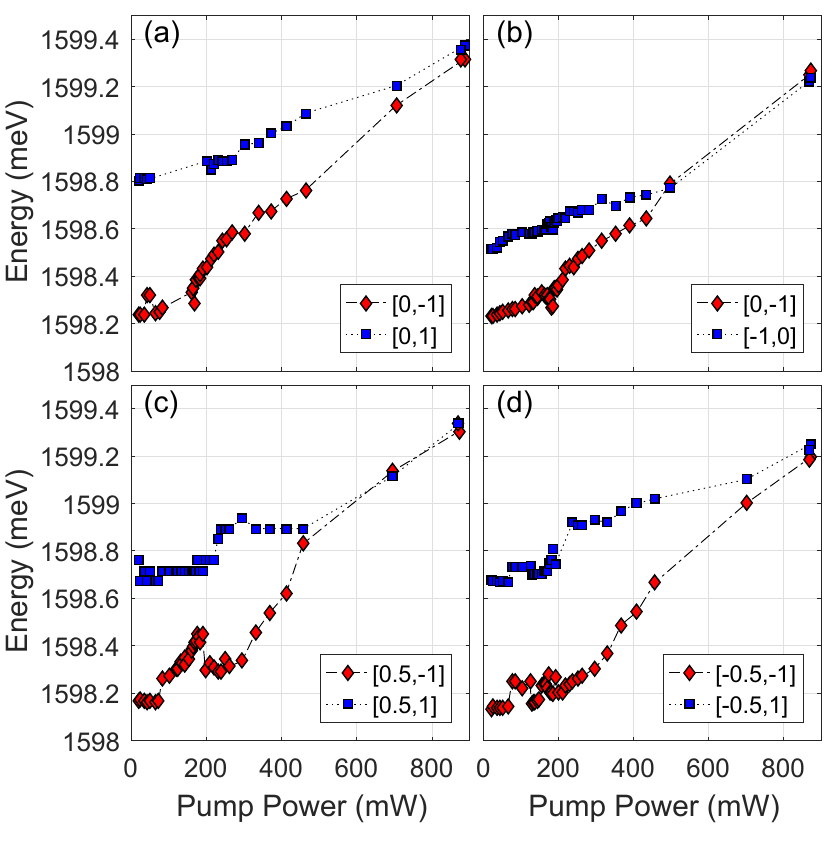}
\caption{Energy vs. pump power at separate locations on the pillar. Data for each pair of locations were taken from a full energy resolved slice across the pillar, similar to and including those in Figure \ref{EvR}, allowing simultaneous measurement of both locations. The locations are defined according to Figure \ref{thresholds}(b) and are (a) [0, -1], [0, 1], (b) [0, -1], [-1, 0], (c) [0.5, -1], [0.5, 1], and (d) [-0.5, -1], [-0.5, 1], shown as red diamonds and blue squares, respectively. The uncertainty is $\pm0.05\:$meV.}
\label{EvP}
\end{figure}

The merging of the separated condensates into one, single condensate can be seen clearly in the data of Figure \ref{EvP}. In this figure we compare the energy vs. pump power at various separated locations on the pillar. The data were taken at two spatial locations simultaneously by imaging a full slice across the pillar. This removes deviations in the measurements due to pump laser pointing instabilities or small shifts of the pump location. For each location, energy blue-shifting was seen above the threshold power, with the lower energy locations showing larger blue shifts due to higher polariton densities. In all cases, the lower energy locations eventually shifted to the same energy as the higher energy locations, and then their energies were locked together above that power to within out spectral uncertainty ($\pm0.05\:$meV). This same feature is apparent in Figure \ref{EvR}(c-d), which correspond to two pairs of data points in Figure \ref{EvP}(a). This indicates that the apparently interconnected polariton luminescence at powers well above threshold shown in Figure \ref{Rspace}(d) was from a single mono-energetic polariton condensate.


Future work on these edge-trapped condensates could include interfering the emission from different regions of the pillar to confirm and characterize the power dependence of a single spatially extended state. Also, since our spectral resolution was insufficient to precisely measure the energy linewidth of these long-lifetime polaritons, it would be worthwhile to perform a time-resolved pulsed-excitation experiment to determine the effect that the described etch methods have on the lifetime. It would also be useful to examine this system's dependence on the pump spot location, which would help with understanding the role of the effective potential of the pump spot on the overall process.

Etch-induced edge trapping could be employed far from corners to produce shallow  ($\approx1\:$meV) 1D wells in relatively wide structures ($>20\:\mu$m). The deeper trapping that emerges at corners could be implemented for shallow harmonic-like traps. The emergence of the energy minima $\approx7\:\mu$m away from the etched edge could allow for even deeper trapping with edges $\leq15\:\mu$m apart without the need for complicated and costly fabrication on the $\sim1\:\mu$m scale.


The work at the University of Pittsburgh was supported by the Army Research Office (W911NF-15-1-0466). The work at Princeton University was supported by the Gordon and Betty Moore Foundation (GBMF-4420) and by the National Science Foundation MRSEC program through the Princeton Center for Complex Materials (DMR-1420541). D.M.M. also acknowledges the support of the Pittsburgh Quantum Institute.

\bibliography{mybib}
\end{document}